\documentclass[aps,prx,twocolumn]{revtex4-2}

\usepackage[normalem]{ulem}
\usepackage{graphicx}
\usepackage{xcolor}
\usepackage{amsmath}
\usepackage{amssymb}
\usepackage{siunitx}
\usepackage[english]{babel}

\usepackage[colorlinks=true, pdfstartview=FitV, linkcolor=blue,
citecolor=blue, urlcolor=blue]{hyperref}
\usepackage[capitalise]{cleveref}

\begin{document}
\title{Imaging short- and long-range magnetic order in a quantum anomalous Hall insulator}
\author{Andriani Vervelaki$^{1}$, Boris Gross$^{1,2}$, Daniel Jetter$^{1}$, Katharina Kress$^{1}$, Timur Weber$^{3}$, Dieter Koelle$^{3}$, Kajetan M. Fijalkowski$^{4,5}$, Martin Klement$^{4,5}$, Nan Liu$^{4,5}$, Karl Brunner$^{4,5}$, Charles Gould$^{4,5}$, Laurens W. Molenkamp$^{4,5}$, Martino Poggio$^{1,2}$, Floris Braakman$^{1,2}$}

\affiliation{$^1$Department of Physics, University of Basel, 4056 Basel, Switzerland,\\ $^2$Swiss Nanoscience Institute, University of Basel, 4056 Basel, Switzerland,\\ $^3$Physikalisches Institut, Center for Quantum Science (CQ) and LISA\textsuperscript{+}, University of T\"ubingen, 72076 T\"ubingen, Germany,\\ $^4$Faculty for Physics and Astronomy (EP3), Universit\"{a}t W\"{u}rzburg, Am Hubland, 97074 W\"{u}rzburg, Germany,\\ $^5$Institute for Topological Insulators, Am Hubland, 97074 W\"{u}rzburg, Germany}

\date{\today}
\begin{abstract}\noindent The quantum anomalous Hall effect has been observed in several magnetically doped topological insulators, where its robustness and macroscopic magnetization properties have been taken to suggest the presence of long-range ferromagnetic order. However, experiments in such systems have found evidence for both long- and short-range order, leaving the precise nature of the magnetism in these systems unclear. Here, we use scanning superconducting quantum interference device microscopy to study magnetic domains in V-doped (Bi,Sb)$_2$Te$_3$ exhibiting a quantum anomalous Hall effect with precise quantization. By imaging stray magnetic fields as a function of applied field, we map the formation and evolution of domains through magnetic reversal. We reconstruct the magnetization configuration underlying the measured stray field and find that magnetic domains and crystallographic grains are of similar size. Moreover, magnetic reversal is found to occur through domain expansion, typical of ferromagnets, rather than through nucleation at random sites. Our measurements thus reveal a coexistence of both local magnetic interactions within crystallographic grains and long-range ferromagnetic coupling between grains. This behavior in V-doped (Bi,Sb)$_2$Te$_3$ is markedly distinct from that previously reported for Cr-doped (Bi,Sb)$_2$Te$_3$. 
\end{abstract}

\maketitle
\section*{Introduction}
The quantum anomalous Hall effect (QAHE) in magnetically doped topological insulators yields near-perfect quantization of the Hall resistance to $h/e^2$ at zero applied magnetic field\cite{yuQuantizedAnomalousHall2010,changExperimentalObservationQuantum2013,changHighprecisionRealizationRobust2015a}. Being defined only in terms of constants of nature, such a quantized resistance is of great use in metrological applications\cite{gotzPrecisionMeasurementQuantized2018,foxPartpermillionQuantizationCurrentinduced2018,okazakiQuantumAnomalousHall2022a,rodenbachUnifiedRealizationElectrical2025,huangQuantumAnomalousHall2025,patelZeroExternalMagnetic2024}. Zero-field quantization with a relative inaccuracy of a few parts per billion has recently been measured in a V-doped (Bi,Sb)$_2$Te$_3$ (VBST) sample\cite{patelZeroExternalMagnetic2024}, at a temperature of \SI{34}{\milli\kelvin}. However, while metrological applications of the QAHE seem within reach, the microscopic origins of the effect, in particular its underlying magnetic interactions, remain unclear. Previous transport and imaging experiments have revealed a variety of magnetic behaviors\cite{huangQuantumAnomalousHall2025}, including indications of superparamagnetism\cite{grauerCoincidenceSuperparamagnetismPerfect2015,lachmanVisualizationSuperparamagneticDynamics2015,lachmanObservationSuperparamagnetismCoexistence2017}, but also long-range ferromagnetic interactions\cite{wangVisualizingFerromagneticDomain2016,wangDirectEvidenceFerromagnetism2018,maVisualizationFerromagneticDomains2023} in Cr-doped BST, V-doped BST, and related compounds. In addition, the appearance of complex magnetic phenomena, including magnetic skyrmions\cite{yasudaGeometricHallEffects2016}, coexistence of surface and bulk ferromagnetism\cite{fijalkowskiCoexistenceSurfaceBulk2020}, Barkhausen-like switching\cite{liuLargeDiscreteJumps2016b}, and macroscopic quantum tunneling of magnetization\cite{fijalkowskiMacroscopicQuantumTunneling2023} has been reported. 
Key open questions are what microscopic magnetic structure underlies such a variety of magnetic phenomena and how this microscopic structure affects topological transport properties.

Here, we perform magnetic microscopy on a VBST sample to map its magnetic order. We focus on two key aspects: the role of crystallographic grains in magnetic domain formation, and the evolution of domains through magnetic reversal. A previous transport study has shown that the magnetization of individual domains can flip through macroscopic tunneling\cite{fijalkowskiMacroscopicQuantumTunneling2023}. That study indicated that the size of magnetic domains is similar to that of topographical grains resulting from crystallographic rotational twinning\cite{tarakinaMicrostructuralCharacterisationBi2Se32013,grauerCoincidenceSuperparamagnetismPerfect2015,winnerleinEpitaxyStructuralProperties2017a}, i.e. of the order of 50 to 100 nm. Grain structure in the VBST layer may result in a magnetic exchange interaction that is stronger within the grains than between them, yielding a spatial magnetic profile correlated to the grain structure, as well as potentially giving rise to superparamagnetic behavior.

For our imaging experiments, we use a nanoscale superconducting quantum interference device (SQUID), patterned at the tip of a force-microscopy cantilever~\cite{wyssMagneticThermalTopographic2022a,weberAdvancedSQUIDonleverScanning2025a}. We measure maps of the magnetic stray field above a VBST sample as a function of applied magnetic field, which allows us to characterize the magnetic hysteresis of the sample. Using a magnetic reconstruction method\cite{broadwayImprovedCurrentDensity2020}, we determine possible magnetization maps in agreement with our observations. 
We find that the dimensions of magnetic domains are similar to those of rotational twin grains, indicating that these crystallographic boundaries influence the magnetic order. 
\begin{figure*}[t]
	\includegraphics[width=0.96\textwidth]{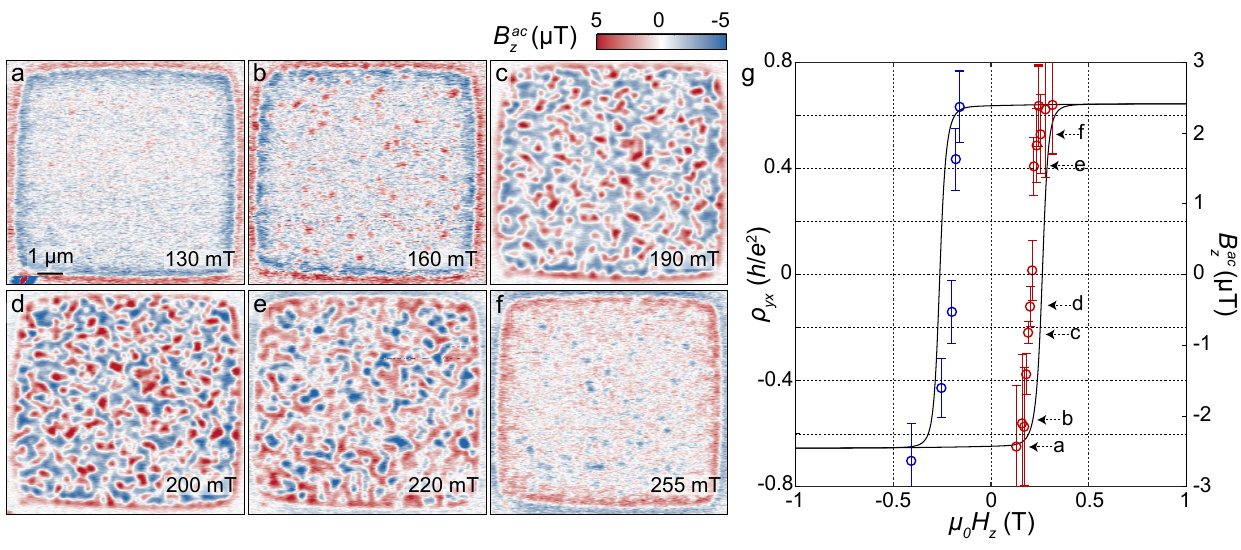}
	\caption{Images of magnetic stray field for various values of the applied magnetic field, covering reversal. \textbf{a - f} $B_z^{ac}$ maps taken at applied magnetic field values spanning one side of the hysteresis loop and showing the magnetic reversal of the sample. Taken with SQUID-sample distance of \SI{260}{\nano\meter}. \textbf{g} Hysteresis loop of the sample (right axis), obtained by averaging $B_z^{ac}$ over an area at one of the edges (see rectangle in Fig.\,S3\textbf{a} for averaging region), as a function of magnetic field. Red (blue) symbols indicate data taken on upward (downward) field sweep. Error bars are calculated from the noise floor of the SQUID read-out at each applied field. Also plotted (solid lines) is a hysteresis loop of the Hall resistivity (left axis), measured at \SI{4.2}{\kelvin} on a Hall bar sample fabricated using the same epilayer as the one measured in this work.
	}
	\label{fig:hyst}
\end{figure*}
Furthermore, we observe that reversal occurs through the expansion (shrinkage) of magnetic domains with magnetization aligned along (opposing) the applied field. 
Such a reversal is typical for long-range ferromagnetic coupling, and was also observed previously in Cr/V co-doped BST using magnetic force microscopy\cite{wangDirectEvidenceFerromagnetism2018}. 
This behavior can be contrasted with indications of superparamagnetism found in Cr-doped BST, where reversal was found to proceed via the nucleation of reversed domains at random sites\cite{lachmanVisualizationSuperparamagneticDynamics2015,lachmanObservationSuperparamagnetismCoexistence2017}, suggesting short-range interactions limited by grain boundaries. Our results exhibit characteristics of both behaviors: we observe domain expansion indicative of long-range ferromagnetic coupling, yet our measured domain sizes are comparable to the grain size, indicating stronger intra-grain than intergrain interactions as would be expected in a superparamagnetic scenario. These observations point to a magnetic ordering of dual nature, where both local grain-boundary constraints and long-range ferromagnetic coupling determine the magnetic state.
\section*{Results}
Our sample consists of a \SI{9}{\nano\meter} thick V$_{0.1}$(Bi$_{0.2}$Sb$_{0.8}$)$_{1.9}$Te$_3$ layer grown by molecular beam epitaxy on a Si(111) substrate and capped in-situ with a \SI{10}{\nano\meter} thick layer of Te\cite{winnerleinEpitaxyStructuralProperties2017a}. The capping layer is necessary to protect the sample surface from degradation due to exposure to lithography and ambient conditions. Transport measurements on a device fabricated from the same epitaxial layer and using the same lithographic process show a clear QAHE with precise quantization at temperatures below \SI{100}{\milli\kelvin} (see Supplemental Material Fig.\,S1). For our magnetic imaging studies, we patterned the sample using Ar-ion milling into a mesa-like structure of roughly \SI{10}{\micro\meter}$\times$\SI{10}{\micro\meter} (see Fig.\,S2 for an optical microscopy image of the mesa studied in this work).
All scanning SQUID microscopy (SSM) measurements presented in this work have been taken at a temperature of \SI{5}{\kelvin}, and at SQUID-sample separations of \SIrange{150}{260}{\nano\meter}. The SQUID sensor has an effective loop diameter of \SI{80}{\nano\meter}, and is fabricated at the apex of an atomic force microscopy (AFM) cantilever. For more details on the SQUID probe, see Weber et al.\cite{weberAdvancedSQUIDonleverScanning2025a}.

\noindent We perform SSM to measure the normal-to-plane stray magnetic field $B_z$ at constant height above the sample. In addition to $B_z$, we also record the quantity $B_z^{ac} \propto dB_z/dz$, which is obtained by sinusoidally modulating the sample's $z$-position at a frequency of \SI{178}{\hertz} and demodulating the SQUID signal at this frequency using a lock-in amplifier. Note that $dB_z/dz$ is proportional to $B_z^{ac}$ but has opposite sign. 
The quantity $B_z^{ac}$ has higher signal-to-noise ratio than $B_z$, due to spectral noise filtering.
\begin{figure*}[t]
	\includegraphics[width=0.9\textwidth]{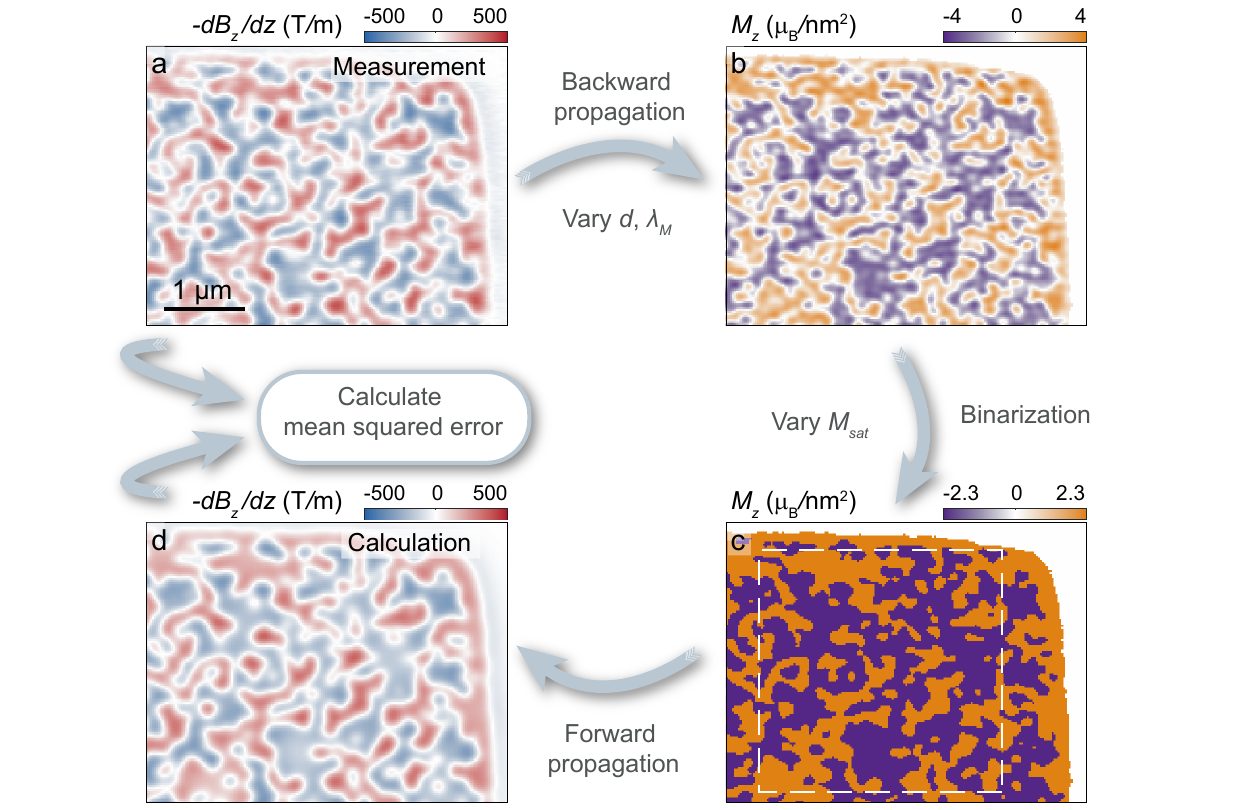}
	\caption{Iterative procedure of reconstructing a magnetization pattern optimally matching the measured stray field at coercive field. \textbf{a} Map of measured $-dB_z/dz$, at $\mathrm{\mu_0} H_z$ = \SI{200}{\milli\tesla} and a calibrated SQUID-sample distance of \SI{150}{\nano\meter}. \textbf{b} Map of reconstructed $M_z$. Note that regions exterior to the mesa have been removed, since any reconstructed magnetization there is an artifact. \textbf{c} Binary $M_z$ map, generated by assigning to positive (negative) values of $M_z$ in the map of \textbf{b} a value of $+M_{sat}$ ($-M_{sat}$). White rectangle indicates area shown in Fig.\,\ref{fig:afm}\textbf{c}. \textbf{d} Map of $-dB_z/dz$ calculated from the binary $M_z$ map in \textbf{c}.\\ 
	By iterating over a range of the parameters $d$, $\lambda_M$, and $M_{sat}$, the configuration minimizing the MSE can be found. The images shown in \textbf{b} - \textbf{d} correspond to this optimal configuration, with $d$ = 157 nm, $\lambda_M$ = 109 nm, and $M_{sat}$ = 2.3 $\mathrm{\mu_B}$/nm$^2$.}
	\label{fig:reconstruction}
\end{figure*}

\noindent We study the magnetic reversal process by imaging $B_z$ and $B_z^{ac}$ as a function of applied magnetic field $\mathrm{\mu_0} H_z$. Figs.\,\ref{fig:hyst}\textbf{a}-\textbf{f} show $B_z^{ac}$ images taken at applied magnetic fields covering magnetic reversal. See Fig.\,S3 for additional $B_z^{ac}$ images taken over an extended range of applied field, with an interval of approximately \SI{10}{\milli\tesla}, as well as Fig.\,S4 for an example $B_z$ image together with the corresponding $B_z^{ac}$ image at \SI{200}{\milli\tesla}. We start the measurements by applying an normal-to-plane magnetic field $\mathrm{\mu_0} H_z$ of \SI{-1}{\tesla}, which results in saturated magnetization. 
Upon increasing the applied field, we find that the stray field profile remains largely unchanged until we reach a reverse applied field of $\mathrm{\mu_0} H_z\approx$ \SI{130}{\milli\tesla}. At this point, we start to see the formation of multiple islands featuring reversed magnetic stray field across the sample, as shown in Fig.\,\ref{fig:hyst}\textbf{b}. 
Additionally, bands of positive and negative stray field appear running along the sample edges. The magnitude of this stray field at the edges is proportional to the integral of the normal-to-plane magnetization of the sample (assuming that the lateral magnetization is negligible, as expected for large normal-to-plane anisotropy).

We proceed by incrementally increasing the applied field and taking images every $\sim$\SI{10}{\milli\tesla}. Both the number of islands with reversed stray field and their area increases with applied field, as can be seen in Figs.\,\ref{fig:hyst}\textbf{c}-\textbf{f}. We find the coercive field, corresponding to zero net magnetization, to be $\sim$\SI{200}{\milli\tesla} (obtained by averaging over the area delineated by a rectangle in Fig.\,S3\textbf{a}). 
At this applied field, there are roughly an equal amount of regions with positive and negative $B_z^{ac}$. Consistently, the stray field around the edges of the sample is minimized at the coercive field.
Increasing the field further leads to continued build-up of reversed stray field, until saturation is reached at $\sim$\SI{300}{\milli\tesla}.

We observe a clear hysteretic behavior of the stray field upon sweeping the applied magnetic field upward or downward from saturation. Fig.\,\ref{fig:hyst}\textbf{g} shows a plot of the net $B_z^{ac}$ (determined again by averaging over the region indicated by the rectangle in Fig.\,S3\textbf{a}), as a function of applied magnetic field. The measurements were obtained for a range of applied fields reached by both up- and downward sweeping. We compare this hysteresis loop with that of the Hall resistivity $\rho_{yx}$ taken at a temperature of \SI{4.2}{\kelvin} with a Hall bar device fabricated from the same epilayer, which shows a coercive field of $\sim$\SI{250}{\milli\tesla}. The coercive field extracted from imaging is consistent with a temperature slightly above \SI{4.2}{\kelvin}.\\ 

To determine whether the observed stray field maps could be produced by magnetic domains related to the crystallographic grains, we follow the procedure sketched in Fig.\,\ref{fig:reconstruction} and detailed in Supplemental Material Section IV. The figure illustrates the procedure using the optimal parameter set. Starting from the image of $B_z^{ac}$ taken at coercive field (Fig.\,\ref{fig:reconstruction}\textbf{a}, converted to $-dB_z/dz$ using the oscillation peak-to-peak amplitude of \SI{50}{\nano\meter}), we reconstruct a map of the normal-to-plane magnetization $M_z$ using a reverse propagation method\cite{broadwayImprovedCurrentDensity2020}. We assume the magnetization is aligned either up or down\cite{changHighprecisionRealizationRobust2015a} and confined to a 2D plane corresponding to the thin VBST layer, resulting in the map shown in Fig.\,\ref{fig:reconstruction}\textbf{b}. This reconstruction depends on the chosen SQUID-sample distance $d$ and Hann filter cut-off wavelength $\lambda_M$.

To verify whether a binary magnetization profile could reproduce our measurements, we threshold the reconstructed map by assigning saturation magnetization values $+M_{sat}$ ($-M_{sat}$) to all positive (negative) values of $M_z(x,y)$, yielding Fig.\,\ref{fig:reconstruction}\textbf{c}. This binary profile, with individual domains magnetized fully up or down, is consistent with strong normal-to-plane anisotropy\cite{changHighprecisionRealizationRobust2015a}. We then calculate $-dB_z/dz$ from this binary map (Fig.\,\ref{fig:reconstruction}\textbf{d}) and compare it to the measurement of Fig.\,\ref{fig:reconstruction}\textbf{a} by calculating the mean squared error (MSE) using pixelwise subtraction. 

Iterating over the parameters $d$, $\lambda_M$, and $M_{sat}$ to minimize MSE yields optimal values $d = 157$ nm, $\lambda_M = 109$ nm, and $M_{sat} = 2.3$ $\mathrm{\mu_B}$/nm$^2$, with 95\% confidence intervals $d = 146$-$171$ nm and $M_{sat} = 2.0$-$2.8$ $\mathrm{\mu_B}$/nm$^2$ (see Fig.\,S5). The agreement between measured and calculated maps for the optimum parameter set validates the binary magnetization profile of Fig.\,\ref{fig:reconstruction}c. Given lattice constants reported in previous works\cite{dyckDilutedMagneticSemiconductors2002a,winnerleinEpitaxyStructuralProperties2017a} and a V doping concentration of 0.1 ions per unit cell, the above values of $M_{sat}$ correspond to a magnetic moment of 1.4 - 1.8 $\mathrm{\mu_B}$ per dopant ion, which overlaps with what has previously been found both experimentally and theoretically for V-dopants in BST\cite{changHighprecisionRealizationRobust2015a,tcakaevComparingMagneticGroundstate2020a}.\\

\begin{figure}[t]
	\includegraphics[width=0.47\textwidth]{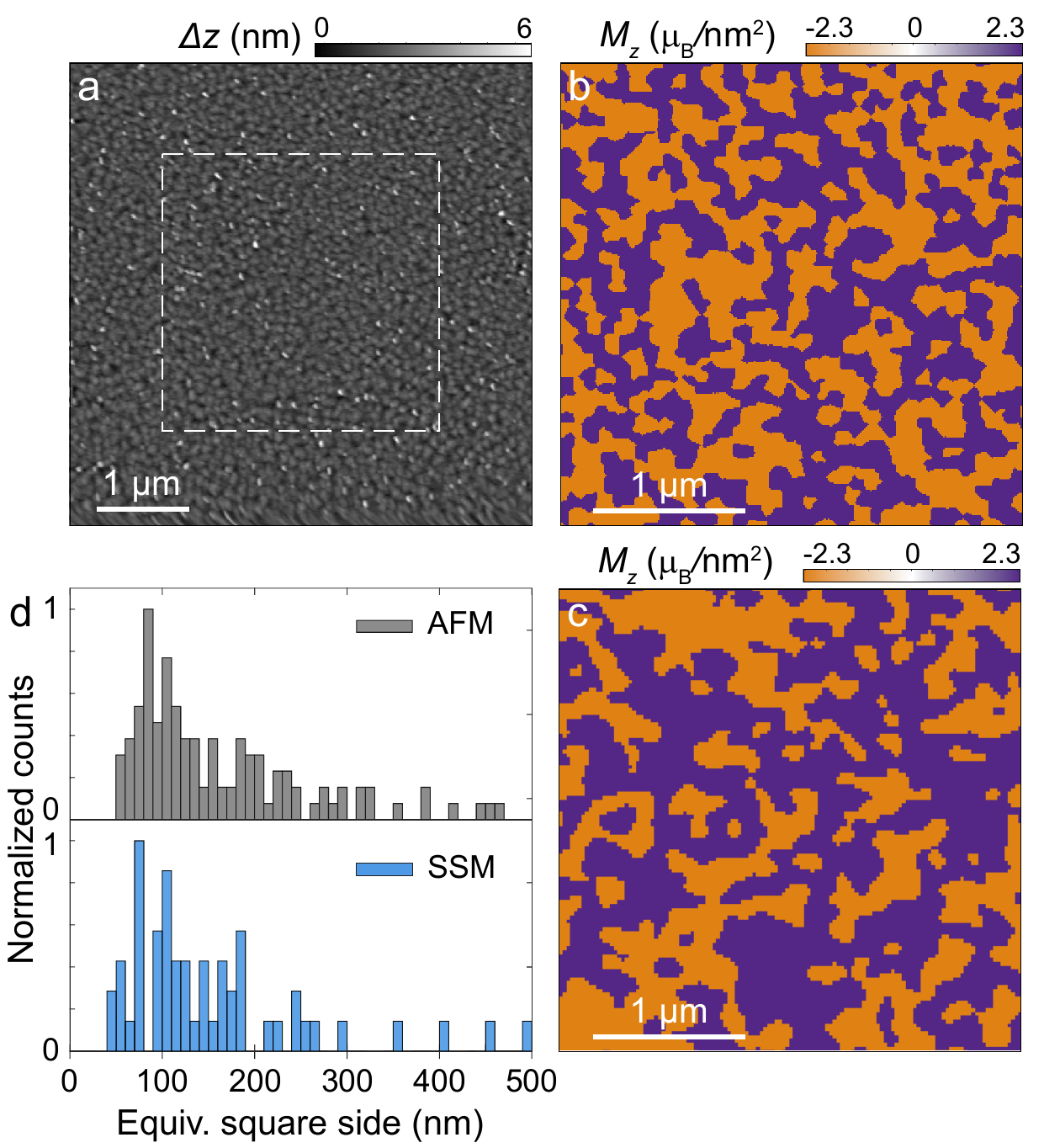}
	\caption{Comparison of topographical structure with reconstructed magnetization. \textbf{a} Height map of an uncapped VBST film, measured through AFM. \textbf{b} Magnetization map, created by randomly assigning values of $\pm 2.3\mathrm{\mu_B}$/nm$^2$ with equal probability to each of the grains in the 3$\times$\SI{3}{\micro\meter\squared} area outlined in \textbf{a}. \textbf{c} 3$\times$\SI{3}{\micro\meter\squared} part of the reconstructed map of Fig.\,\ref{fig:reconstruction}\textbf{c}. \textbf{d} Comparison of normalized histograms of equivalent square side sizes of domains in the magnetization map Fig.\,\ref{fig:afm}\textbf{b} (upper panel) and of domains in the reconstructed magnetization map of Fig.\,\ref{fig:reconstruction}\textbf{c} (lower panel). Note that sizes larger than 500 nm are not plotted.
	}
	\label{fig:afm}
\end{figure}

Armed with this magnetization map, we compare the magnetization configuration of Fig.\,\ref{fig:reconstruction}\textbf{c} to a plausible magnetization configuration generated by assigning to each topographical grain an normal-to-plane magnetization of equal magnitude but with random sign. We analyze the grain profile found in an AFM image (Fig.\,\ref{fig:afm}\textbf{a}) of a reference VBST layer grown under nominally identical conditions, but without the Te capping. The use of a separate uncapped layer is necessary, as the amorphous Te cap on the VBST layer used for magnetic imaging obscures the VBST crystal grain height profile. First, we segment the height map of the AFM measurement into individual grains using a thresholding algorithm. Next, we randomly assign a magnetization to each of the grains of either $-2.3\,\mathrm{\mu_B}$/nm$^2$ or $+2.3\,\mathrm{\mu_B}$/nm$^2$, with equal probability. This results in the binary magnetization map of Fig.\,\ref{fig:afm}\textbf{b}. Comparing this to a similarly sized area of the reconstructed magnetization map of Fig.\,\ref{fig:reconstruction}\textbf{c}, shown in Fig.\,\ref{fig:afm}\textbf{c}, we find qualitative agreement. More examples of a random magnetization direction assignment to the crystal grains can be found in Fig.\,S6.\\

We analyze and compare the distribution of magnetic domain sizes in the maps of Fig.\,\ref{fig:reconstruction}\textbf{c} and Fig.\,\ref{fig:afm}\textbf{b}, using a watershed-based segmentation algorithm. Fig.\,\ref{fig:afm}\textbf{d} shows two histograms of the equivalent square side sizes of the domains identified in both maps. We observe agreement between the distribution of domain sizes in the reconstructed magnetization map of Fig.\,\ref{fig:reconstruction}\textbf{c} and in the randomly assigned magnetization map of Fig.\,\ref{fig:afm}\textbf{b}. Both distributions peak at $\sim$85\,-\,100\,nm and fall off rapidly below this value (where our finite sensor size is also limiting resolution), whereas there is a longer tail on the high side, extending approximately to \SI{300}{\nano\meter}. This size distribution is consistent with that of crystallographic rotational twin grain sizes in VBST\cite{winnerleinEpitaxyStructuralProperties2017a,fijalkowskiMacroscopicQuantumTunneling2023} and in the similar compound Bi$_2$Se$_3$\cite{tarakinaMicrostructuralCharacterisationBi2Se32013}, providing further evidence that the formation of magnetic domains in VBST is correlated with crystal grain structure. 
Note that the domain-size comparison above is restricted to the sub-500\,nm range relevant for direct correlation with crystallographic grain size. A visual inspection of Figs.\,\ref{fig:afm}\textbf{b} and \textbf{c} reveals, however, that the latter contains a markedly higher proportion of larger domains. This feature points to non-negligible intergrain magnetic interactions, whereby magnetization reversal preferentially occurs in grains adjacent to already-reversed ones.

\begin{figure}[t]
	\includegraphics[width=0.47\textwidth]{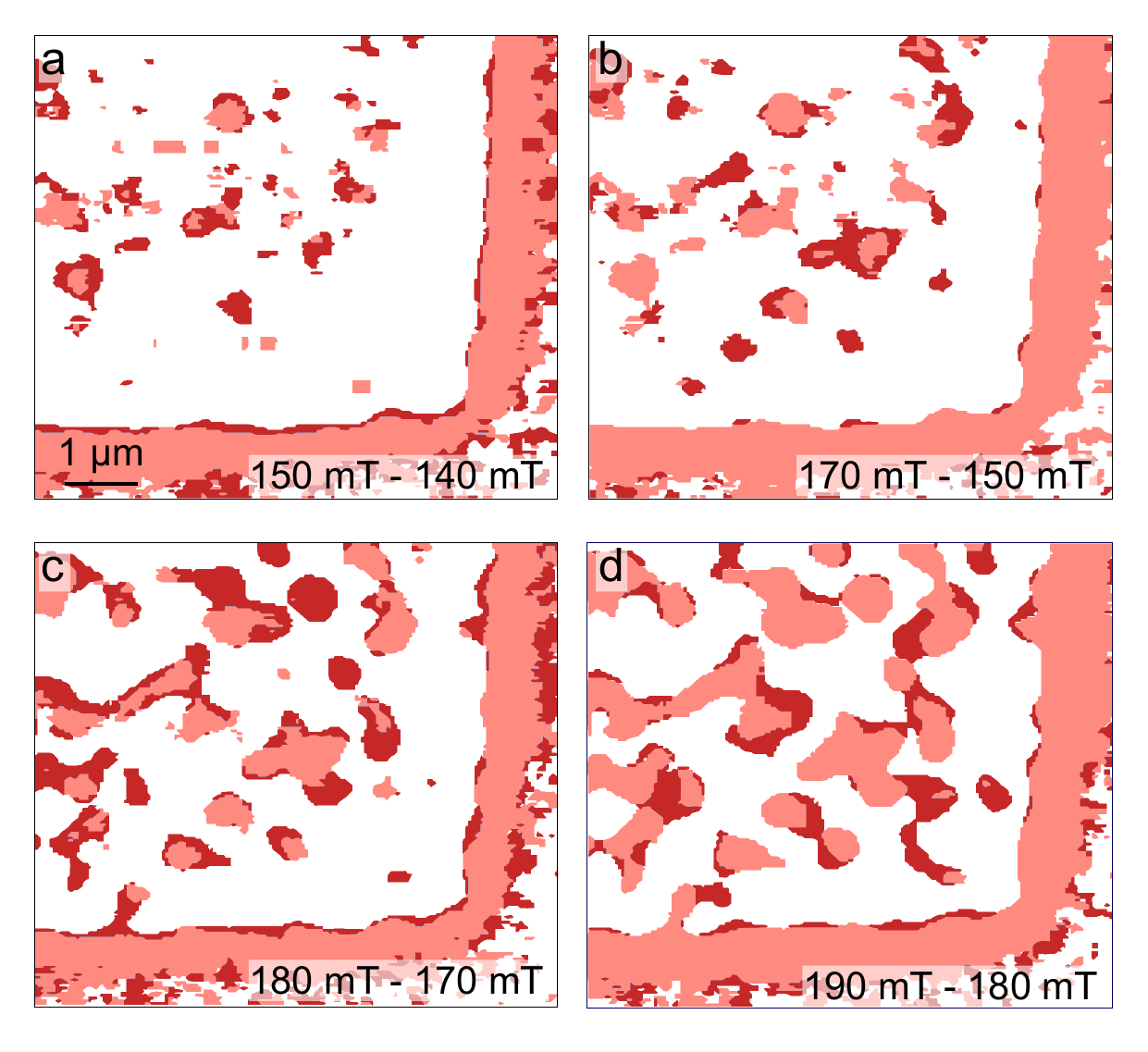}
	\caption{Differential images of measured $B_z^{ac}$ showing magnetic domain reversal progression between consecutive applied fields (data from Fig.\,S3\textbf{c-d},\textbf{f-h}). Dark red: newly reversed regions that flipped between the two indicated field values. Light red: regions that had already reversed at the lower indicated field value. 
	}
	\label{fig:diff}
\end{figure}

To probe such intergrain interactions more directly, we now investigate the evolution of the magnetic domains as the system proceeds through reversal. We visualize this evolution as a function of applied magnetic field by using differential images, obtained by subtracting measurements taken at consecutive applied fields. These images reveal how the reversal evolves spatially across the sample as the field is increased incrementally. To highlight the reversal toward positive magnetization, we only show positive differences here, corresponding to magnetization flipping from being anti-aligned with the applied magnetic field to being aligned with it. Fig.\,\ref{fig:diff} shows such differential images (in dark red), overlaid on top of images of the areas which already showed a positive stray field in the measurement taken at the lower field (shown in light red). Such combined images are shown for four pairs of consecutive applied magnetic fields in Figs.\,\ref{fig:diff} (see Fig.\,S7 for additional differential images). As can be seen, reversal happens predominantly along the edges of previously reversed domains, indicating a gradual expansion of the reversed regions. We see this behavior over the complete range of applied fields covering reversal from negative to positive magnetization.

The behavior observed in Figs.\,\ref{fig:diff} is very different from that seen in imaging experiments on Cr-doped BST\cite{lachmanVisualizationSuperparamagneticDynamics2015,lachmanObservationSuperparamagnetismCoexistence2017}, where reversal was found to proceed via the random switching of local moments. Such random switching indicates that magnetic domains in Cr-doped BST are only weakly coupled across local boundaries, such as topographical grain boundaries or magnetic dopant clusters, resulting in a superparamagnetic state. In our measurements, although we find the magnetic domain size to be correlated with the crystallographic grain size, the observed evolution of domains with applied magnetic field indicates that ferromagnetic coupling is significant on length scales larger than that given by the grain profile (Wang et al.\cite{wangDirectEvidenceFerromagnetism2018} similarly report long-range ferromagnetic coupling in the related Cr/V co-doped BST). This suggests that in VBST, in contrast to the Cr-doped material, magnetic exchange interactions across topographical grain boundaries remain significant. 

We further investigate the domain growth behavior by performing micromagnetic simulations using MuMax3\cite{vansteenkisteDesignVerificationMuMax32014,exlLaBontesMethodRevisited2014,leliaertCurrentdrivenDomainWall2014,leliaertAdaptivelyTimeStepping2017a}. We consider various values of the intergrain magnetic exchange stiffness $A_{ex,inter}$ and find three distinct reversal regimes. For $A_{ex,inter} \lesssim 5\times 10^{-16} \mathrm{J\,m^{-1}}$, predominantly independent magnetization switching of individual grains occurs. This regime resembles the superparamagnetic reversal found by Lachman et al.\cite{lachmanVisualizationSuperparamagneticDynamics2015,lachmanObservationSuperparamagnetismCoexistence2017}. In contrast, for $A_{ex,inter} \geq 10^{-14} \mathrm{J\,m^{-1}}$, switching occurs simultaneously across the entire system. Such values of $A_{ex,inter}$ approach the intragrain exchange stiffness $A_{ex,intra}$, which is likely in the range from $10^{-14}$ to $10^{-11} \mathrm{J\,m^{-1}}$. In the intermediate regime with $A_{ex,inter}$ in the range $10^{-15} - 10^{-14} \mathrm{J\,m^{-1}}$, we reproduce the behavior found in Fig.\,\ref{fig:diff}, where cooperative grain switching leads to domains growing along their edges through reversal. See Supplemental Material Section VII for details and Supplemental Material Movie I for an illustration of the simulated domain growth.

\section*{Conclusions and outlook}
We have characterized the magnetic structure of VBST using SSM. The observed magnetic domain sizes correspond well to the crystallographic grain sizes resulting from rotational twinning in the material. Crystal grain boundaries locally perturb exchange coupling and magnetocrystalline anisotropy, making them natural pinning sites for magnetic domain walls.
Magnetic images acquired over a range of applied fields yield a hysteresis loop of the stray magnetic field that agrees well with corresponding transport measurements. We find that the domains evolve as a function of applied magnetic field through growth along their edges.
These observations highlight that magnetism in VBST is a result of an interplay between short-range magnetic interactions which accommodate the local magnetic profile to that of the crystal landscape, and long-range ferromagnetic coupling that stabilizes the global magnetic alignment. Any comprehensive model attempting to explain magnetism in this material needs to incorporate both aspects.

Note that transport measurements taken between \SI{40}{\milli\kelvin} and \SI{440}{\milli\kelvin} show evidence for thermally activated magnetic switching of individual domains in similar samples\cite{fijalkowskiMacroscopicQuantumTunneling2023}. Given that the measurements shown here are taken at \SI{5}{\kelvin}, such switching likely occurs on timescales so short that our SSM measurements average over such effects. It would, however, be of interest to perform SSM at lower temperatures, where these switching events become slow enough to be observable, and to map their spatial dependence. Also, imaging the magnetic configuration as well as the current flow through a Hall bar made out of VBST could give insight into the role of the magnetic domain structure in determining the spatial profile of edge channels, such as by following a Landau-B\"uttiker network of 1D channels around the magnetic domains.

\section*{Acknowledgements}
We acknowledge support from the European Commision under H2020 FET Open
grant “FIBsuperProbes” (Grant No. 892427), as well as through the Project 23FUN07 QuAHMET,
which has received funding from the European Partnership on
Metrology, co-financed by the European Union’s Horizon Europe
Research and Innovation Programme and by the participating states. We further acknowledge support from the Free State of Bavaria (Institute for Topological Insulators), the German Research Foundation (Project SFB 1170, Grant No. 258499086), and the W\"urzburg-Dresden Cluster of Excellence on Complexity, Topology and Dynamics in Quantum Matter (EXC 2147, 390858490).

\newpage

\bibliography{VBSTv5}

@article{broadwayImprovedCurrentDensity2020,
  title = {Improved {{Current Density}} and {{Magnetization Reconstruction Through Vector Magnetic Field Measurements}}},
  author = {Broadway, D.A. and Lillie, S.E. and Scholten, S.C. and Rohner, D. and Dontschuk, N. and Maletinsky, P. and Tetienne, J.-P. and Hollenberg, L.C.L.},
  year = 2020,
  month = aug,
  journal = {Phys. Rev. Appl.},
  volume = {14},
  number = {2},
  pages = {024076},
  publisher = {American Physical Society},
  doi = {10.1103/PhysRevApplied.14.024076},
  urldate = {2024-06-14}
}

@article{changExperimentalObservationQuantum2013,
  title = {Experimental {{Observation}} of the {{Quantum Anomalous Hall Effect}} in a {{Magnetic Topological Insulator}}},
  author = {Chang, Cui-Zu and Zhang, Jinsong and Feng, Xiao and Shen, Jie and Zhang, Zuocheng and Guo, Minghua and Li, Kang and Ou, Yunbo and Wei, Pang and Wang, Li-Li and Ji, Zhong-Qing and Feng, Yang and Ji, Shuaihua and Chen, Xi and Jia, Jinfeng and Dai, Xi and Fang, Zhong and Zhang, Shou-Cheng and He, Ke and Wang, Yayu and Lu, Li and Ma, Xu-Cun and Xue, Qi-Kun},
  year = 2013,
  month = apr,
  journal = {Science},
  volume = {340},
  number = {6129},
  pages = {167--170},
  publisher = {American Association for the Advancement of Science},
  doi = {10.1126/science.1234414},
  urldate = {2024-04-30}
}

@article{changHighprecisionRealizationRobust2015a,
  title = {High-Precision Realization of Robust Quantum Anomalous {{Hall}} State in a Hard Ferromagnetic Topological Insulator},
  author = {Chang, Cui-Zu and Zhao, Weiwei and Kim, Duk Y. and Zhang, Haijun and Assaf, Badih A. and Heiman, Don and Zhang, Shou-Cheng and Liu, Chaoxing and Chan, Moses H. W. and Moodera, Jagadeesh S.},
  year = 2015,
  month = may,
  journal = {Nature Mater},
  volume = {14},
  number = {5},
  pages = {473--477},
  publisher = {Nature Publishing Group},
  issn = {1476-4660},
  doi = {10.1038/nmat4204},
  urldate = {2024-06-26},
  copyright = {2014 Springer Nature Limited},
  langid = {english}
}

@article{dyckDilutedMagneticSemiconductors2002a,
  title = {Diluted magnetic semiconductors based on {Sb}$_{2-x}${V}$_{x}${Te}$_{3}$ ($0.01 \lesssim x \lesssim 0.03$)},
  shorttitle = {Diluted magnetic semiconductors based on <span class="aps-inline-formula"><math xmlns="http},
  author = {Dyck, Jeffrey S.},
  year = 2002,
  journal = {Phys. Rev. B},
  volume = {65},
  number = {11},
  pages = {115212},
  doi = {10.1103/PhysRevB.65.115212}
}

@article{exlLaBontesMethodRevisited2014,
  title = {{{LaBonte}}'s Method Revisited: {{An}} Effective Steepest Descent Method for Micromagnetic Energy Minimization},
  shorttitle = {{{LaBonte}}'s Method Revisited},
  author = {Exl, Lukas and Bance, Simon and Reichel, Franz and Schrefl, Thomas and Peter Stimming, Hans and Mauser, Norbert J.},
  year = 2014,
  month = jan,
  journal = {Journal of Applied Physics},
  volume = {115},
  number = {17},
  pages = {17D118},
  issn = {0021-8979},
  doi = {10.1063/1.4862839},
  urldate = {2023-06-08}
}

@article{fijalkowskiCoexistenceSurfaceBulk2020,
  title = {Coexistence of {{Surface}} and {{Bulk Ferromagnetism Mimics Skyrmion Hall Effect}} in a {{Topological Insulator}}},
  author = {Fijalkowski, K. M. and Hartl, M. and Winnerlein, M. and Mandal, P. and Schreyeck, S. and Brunner, K. and Gould, C. and Molenkamp, L. W.},
  year = 2020,
  month = jan,
  journal = {Phys. Rev. X},
  volume = {10},
  number = {1},
  pages = {011012},
  publisher = {American Physical Society},
  doi = {10.1103/PhysRevX.10.011012},
  urldate = {2024-05-03}
}

@article{fijalkowskiMacroscopicQuantumTunneling2023,
  title = {Macroscopic {{Quantum Tunneling}} of a {{Topological Ferromagnet}}},
  author = {Fijalkowski, Kajetan M. and Liu, Nan and Mandal, Pankaj and Schreyeck, Steffen and Brunner, Karl and Gould, Charles and Molenkamp, Laurens W.},
  year = 2023,
  journal = {Advanced Science},
  volume = {10},
  number = {22},
  pages = {2303165},
  issn = {2198-3844},
  doi = {10.1002/advs.202303165},
  urldate = {2024-06-21},
  langid = {english}
}

@article{foxPartpermillionQuantizationCurrentinduced2018,
  title = {Part-per-Million Quantization and Current-Induced Breakdown of the Quantum Anomalous {{Hall}} Effect},
  author = {Fox, E. J. and Rosen, I. T. and Yang, Yanfei and Jones, George R. and Elmquist, Randolph E. and Kou, Xufeng and Pan, Lei and Wang, Kang L. and {Goldhaber-Gordon}, D.},
  year = 2018,
  month = aug,
  journal = {Phys. Rev. B},
  volume = {98},
  number = {7},
  pages = {075145},
  publisher = {American Physical Society},
  doi = {10.1103/PhysRevB.98.075145},
  urldate = {2025-01-03}
}

@article{gotzPrecisionMeasurementQuantized2018,
  title = {Precision Measurement of the Quantized Anomalous {{Hall}} Resistance at Zero Magnetic Field},
  author = {G{\"o}tz, Martin and Fijalkowski, Kajetan M. and Pesel, Eckart and Hartl, Matthias and Schreyeck, Steffen and Winnerlein, Martin and Grauer, Stefan and Scherer, Hansj{\"o}rg and Brunner, Karl and Gould, Charles and Ahlers, Franz J. and Molenkamp, Laurens W.},
  year = 2018,
  month = feb,
  journal = {Applied Physics Letters},
  volume = {112},
  number = {7},
  pages = {072102},
  issn = {0003-6951},
  doi = {10.1063/1.5009718},
  urldate = {2024-08-14}
}

@article{grauerCoincidenceSuperparamagnetismPerfect2015,
  title = {Coincidence of Superparamagnetism and Perfect Quantization in the Quantum Anomalous {{Hall}} State},
  author = {Grauer, S. and Schreyeck, S. and Winnerlein, M. and Brunner, K. and Gould, C. and Molenkamp, L. W.},
  year = 2015,
  month = nov,
  journal = {Phys. Rev. B},
  volume = {92},
  number = {20},
  pages = {201304},
  publisher = {American Physical Society},
  doi = {10.1103/PhysRevB.92.201304},
  urldate = {2024-07-01}
}

@article{huangQuantumAnomalousHall2025,
  title = {Quantum Anomalous {{Hall}} Effect for Metrology},
  author = {Hu{\'a}ng, Nathaniel J. and Boland, Jessica L. and Fijalkowski, Kajetan M. and Gould, Charles and Hesjedal, Thorsten and Kazakova, Olga and Kumar, Susmit and Scherer, Hansj{\"o}rg},
  year = 2025,
  month = jan,
  journal = {Applied Physics Letters},
  volume = {126},
  number = {4},
  pages = {040501},
  issn = {0003-6951},
  doi = {10.1063/5.0233689},
  urldate = {2025-06-03}
}

@article{lachmanObservationSuperparamagnetismCoexistence2017,
  title = {Observation of Superparamagnetism in Coexistence with Quantum Anomalous {{Hall C}} = \textpm 1 and {{C}} = 0 {{Chern}} States},
  author = {Lachman, Ella O. and Mogi, Masataka and Sarkar, Jayanta and Uri, Aviram and Bagani, Kousik and Anahory, Yonathan and Myasoedov, Yuri and Huber, Martin E. and Tsukazaki, Atsushi and Kawasaki, Masashi and Tokura, Yoshinori and Zeldov, Eli},
  year = 2017,
  month = dec,
  journal = {npj Quant Mater},
  volume = {2},
  number = {1},
  pages = {1--7},
  publisher = {Nature Publishing Group},
  issn = {2397-4648},
  doi = {10.1038/s41535-017-0072-1},
  urldate = {2024-07-11},
  copyright = {2017 The Author(s)},
  langid = {english}
}

@article{lachmanVisualizationSuperparamagneticDynamics2015,
  title = {Visualization of Superparamagnetic Dynamics in Magnetic Topological Insulators},
  author = {Lachman, Ella O. and Young, Andrea F. and Richardella, Anthony and Cuppens, Jo and Naren, H. R. and Anahory, Yonathan and Meltzer, Alexander Y. and Kandala, Abhinav and Kempinger, Susan and Myasoedov, Yuri and Huber, Martin E. and Samarth, Nitin and Zeldov, Eli},
  year = 2015,
  month = nov,
  journal = {Science Advances},
  volume = {1},
  number = {10},
  pages = {e1500740},
  publisher = {American Association for the Advancement of Science},
  doi = {10.1126/sciadv.1500740},
  urldate = {2024-06-26}
}

@article{leliaertAdaptivelyTimeStepping2017a,
  title = {Adaptively Time Stepping the Stochastic {{Landau-Lifshitz-Gilbert}} Equation at Nonzero Temperature: {{Implementation}} and Validation in {{MuMax3}}},
  shorttitle = {Adaptively Time Stepping the Stochastic {{Landau-Lifshitz-Gilbert}} Equation at Nonzero Temperature},
  author = {Leliaert, J. and Mulkers, J. and De Clercq, J. and Coene, A. and Dvornik, M. and Van Waeyenberge, B.},
  year = 2017,
  month = dec,
  journal = {AIP Advances},
  volume = {7},
  number = {12},
  pages = {125010},
  issn = {2158-3226},
  doi = {10.1063/1.5003957},
  urldate = {2026-01-22}
}

@article{leliaertCurrentdrivenDomainWall2014,
  title = {Current-Driven Domain Wall Mobility in Polycrystalline Permalloy Nanowires: {{A}} Numerical Study},
  shorttitle = {Current-Driven Domain Wall Mobility in Polycrystalline {{Permalloy}} Nanowires},
  author = {Leliaert, J. and {Van de Wiele}, B. and Vansteenkiste, A. and Laurson, L. and Durin, G. and Dupr{\'e}, L. and Van Waeyenberge, B.},
  year = 2014,
  month = jun,
  journal = {J. Appl. Phys.},
  volume = {115},
  number = {23},
  pages = {233903},
  issn = {0021-8979},
  doi = {10.1063/1.4883297},
  urldate = {2026-01-22}
}

@article{liuLargeDiscreteJumps2016b,
  title = {Large Discrete Jumps Observed in the Transition between {{Chern}} States in a Ferromagnetic Topological Insulator},
  author = {Liu, Minhao and Wang, Wudi and Richardella, Anthony R. and Kandala, Abhinav and Li, Jian and Yazdani, Ali and Samarth, Nitin and Ong, N. Phuan},
  year = 2016,
  month = jul,
  journal = {Science Advances},
  volume = {2},
  number = {7},
  pages = {e1600167},
  publisher = {American Association for the Advancement of Science},
  doi = {10.1126/sciadv.1600167},
  urldate = {2024-11-26}
}

@article{maVisualizationFerromagneticDomains2023,
  title = {Visualization of Ferromagnetic Domains in Vanadium-Doped Topological Insulator Thin Films and Heterostructures},
  author = {Ma, Ying-Jie and Xia, Ti-Rui and Wang, Wen-Bo},
  year = 2023,
  month = sep,
  journal = {Tungsten},
  volume = {5},
  number = {3},
  pages = {288--299},
  issn = {2661-8036},
  doi = {10.1007/s42864-022-00140-x},
  urldate = {2024-07-09},
  langid = {english}
}

@article{okazakiQuantumAnomalousHall2022a,
  title = {Quantum Anomalous {{Hall}} Effect with a Permanent Magnet Defines a Quantum Resistance Standard},
  author = {Okazaki, Yuma and Oe, Takehiko and Kawamura, Minoru and Yoshimi, Ryutaro and Nakamura, Shuji and Takada, Shintaro and Mogi, Masataka and Takahashi, Kei S. and Tsukazaki, Atsushi and Kawasaki, Masashi and Tokura, Yoshinori and Kaneko, Nobu-Hisa},
  year = 2022,
  month = jan,
  journal = {Nat. Phys.},
  volume = {18},
  number = {1},
  pages = {25--29},
  publisher = {Nature Publishing Group},
  issn = {1745-2481},
  doi = {10.1038/s41567-021-01424-8},
  urldate = {2025-07-07},
  copyright = {2021 The Author(s), under exclusive licence to Springer Nature Limited},
  langid = {english}
}

@article{patelZeroExternalMagnetic2024,
  title = {A zero external magnetic field quantum standard of resistance at the {$10^{-9}$} level},
  author = {Patel, D. K. and Fijalkowski, K. M. and Kruskopf, M. and Liu, N. and Götz, M. and Pesel, E. and Jaime, M. and Klement, M. and Schreyeck, S. and Brunner, K. and Gould, C. and Molenkamp, L. W. and Scherer, H.},
  year = 2024,
  month = dec,
  journal = {Nat Electron},
  volume = {7},
  number = {12},
  pages = {1111-1116},
  publisher = {Nature Publishing Group},
  issn = {2520-1131},
  doi = {10.1038/s41928-024-01295-w},
  urldate = {2025-01-08},
  copyright = {2024 The Author(s), under exclusive licence to Springer Nature Limited},
  langid = {english}
}

@article{rodenbachUnifiedRealizationElectrical2025,
  title = {A Unified Realization of Electrical Quantities from the Quantum {{International System}} of {{Units}}},
  author = {Rodenbach, Linsey K. and Underwood, Jason M. and Tran, Ngoc Thanh Mai and Panna, Alireza R. and Andersen, Molly P. and Barcikowski, Zachary S. and Payagala, Shamith U. and Zhang, Peng and Tai, Lixuan and Wang, Kang L. and Jarrett, Dean G. and Elmquist, Randolph E. and Newell, David B. and Rigosi, Albert F. and {Goldhaber-Gordon}, David},
  year = 2025,
  month = aug,
  journal = {Nat Electron},
  volume = {8},
  number = {8},
  pages = {663--671},
  publisher = {Nature Publishing Group},
  issn = {2520-1131},
  doi = {10.1038/s41928-025-01421-2},
  urldate = {2025-11-04},
  copyright = {2025 This is a U.S. Government work and not under copyright protection in the US; foreign copyright protection may apply},
  langid = {english}
}

@article{tarakinaMicrostructuralCharacterisationBi2Se32013,
  title = {Microstructural characterisation of {Bi$_{2}$Se$_{3}$} thin films},
  author = {Tarakina, N V and Schreyeck, S and Borzenko, T and Grauer, S and Schumacher, C and Karczewski, G and Gould, C and Brunner, K and Buhmann, H and Molenkamp, L W},
  year = 2013,
  month = nov,
  journal = {J. Phys.: Conf. Ser.},
  volume = {471},
  number = {1},
  pages = {012043},
  issn = {1742-6596},
  doi = {10.1088/1742-6596/471/1/012043},
  urldate = {2025-02-05},
  langid = {english}
}

@article{tcakaevComparingMagneticGroundstate2020a,
  title = {Comparing magnetic ground-state properties of the V- and Cr-doped topological insulator {(Bi,Sb)}$_{2}${Te}$_{3}$},
  author = {Tcakaev, A. and Zabolotnyy, V. B. and Green, R. J. and Peixoto, T. R. F. and Stier, F. and Dettbarn, M. and Schreyeck, S. and Winnerlein, M. and Vidal, R. Crespo and Schatz, S. and Vasili, H. B. and Valvidares, M. and Brunner, K. and Gould, C. and Bentmann, H. and Reinert, F. and Molenkamp, L. W. and Hinkov, V.},
  year = 2020,
  month = jan,
  journal = {Phys. Rev. B},
  volume = {101},
  number = {4},
  pages = {045127},
  publisher = {American Physical Society},
  doi = {10.1103/PhysRevB.101.045127},
  urldate = {2025-10-23}
}

@article{vansteenkisteDesignVerificationMuMax32014,
  ids = {vansteenkiste_design_2014-1},
  title = {The Design and Verification of {{MuMax3}}},
  author = {Vansteenkiste, Arne and Leliaert, Jonathan and Dvornik, Mykola and Helsen, Mathias and {Garcia-Sanchez}, Felipe and Van Waeyenberge, Bartel},
  year = 2014,
  month = oct,
  journal = {AIP Advances},
  volume = {4},
  number = {10},
  pages = {107133},
  publisher = {American Institute of Physics},
  issn = {2158-3226},
  doi = {10.1063/1.4899186},
  urldate = {2016-09-20},
  langid = {english}
}

@article{wangDirectEvidenceFerromagnetism2018,
  title = {Direct Evidence of Ferromagnetism in a Quantum Anomalous {{Hall}} System},
  author = {Wang, Wenbo and Ou, Yunbo and Liu, Chang and Wang, Yayu and He, Ke and Xue, Qi-Kun and Wu, Weida},
  year = 2018,
  month = aug,
  journal = {Nature Phys},
  volume = {14},
  number = {8},
  pages = {791--795},
  publisher = {Nature Publishing Group},
  issn = {1745-2481},
  doi = {10.1038/s41567-018-0149-1},
  urldate = {2024-11-06},
  copyright = {2018 The Author(s)},
  langid = {english}
}

@article{wangVisualizingFerromagneticDomain2016,
  title = {Visualizing Ferromagnetic Domain Behavior of Magnetic Topological Insulator Thin Films},
  author = {Wang, Wenbo and Chang, Cui-Zu and Moodera, Jagadeesh S. and Wu, Weida},
  year = 2016,
  month = oct,
  journal = {npj Quant Mater},
  volume = {1},
  number = {1},
  pages = {1--5},
  publisher = {Nature Publishing Group},
  issn = {2397-4648},
  doi = {10.1038/npjquantmats.2016.23},
  urldate = {2024-07-11},
  copyright = {2016 The Author(s)},
  langid = {english}
}

@article{weberAdvancedSQUIDonleverScanning2025a,
  title = {Advanced {{SQUID-on-lever}} Scanning Probe for High-Sensitivity Magnetic Microscopy with Sub-100-Nm Spatial Resolution},
  author = {Weber, Timur and Jetter, Daniel and Ullmann, Jan and Koch, Simon A. and Pfander, Simon F. and Kress, Katharina and Vervelaki, Andriani and Gross, Boris and Kieler, Oliver and Drechsler, Ute and Baral, Priya R. and Magrez, Arnaud and Kleiner, Reinhold and Knoll, Armin W. and Poggio, Martino and Koelle, Dieter},
  year = 2025,
  month = nov,
  journal = {Phys. Rev. Appl.},
  volume = {24},
  number = {5},
  pages = {054041},
  publisher = {American Physical Society},
  doi = {10.1103/6s24-vz3k},
  urldate = {2025-11-21}
}

@article{winnerleinEpitaxyStructuralProperties2017a,
  title = {Epitaxy and structural properties of {(V,Bi,Sb)}$_{2}${Te}$_{3}$ layers exhibiting the quantum anomalous {Hall} effect},
  author = {Winnerlein, M. and Schreyeck, S. and Grauer, S. and Rosenberger, S. and Fijalkowski, K. M. and Gould, C. and Brunner, K. and Molenkamp, L. W.},
  year = 2017,
  month = jun,
  journal = {Phys. Rev. Mater.},
  volume = {1},
  number = {1},
  pages = {011201},
  publisher = {American Physical Society},
  doi = {10.1103/PhysRevMaterials.1.011201},
  urldate = {2025-06-04}
}

@article{wyssMagneticThermalTopographic2022a,
  title = {Magnetic, {{Thermal}}, and {{Topographic Imaging}} with a {{Nanometer-Scale SQUID-On-Lever Scanning Probe}}},
  author = {Wyss, M. and Bagani, K. and Jetter, D. and Marchiori, E. and Vervelaki, A. and Gross, B. and Ridderbos, J. and Gliga, S. and Sch{\"o}nenberger, C. and Poggio, M.},
  year = 2022,
  month = mar,
  journal = {Phys. Rev. Applied},
  volume = {17},
  number = {3},
  pages = {034002},
  publisher = {American Physical Society},
  doi = {10.1103/PhysRevApplied.17.034002},
  urldate = {2022-03-02}
}

@article{yasudaGeometricHallEffects2016,
  title = {Geometric {{Hall}} Effects in Topological Insulator Heterostructures},
  author = {Yasuda, K. and Wakatsuki, R. and Morimoto, T. and Yoshimi, R. and Tsukazaki, A. and Takahashi, K. S. and Ezawa, M. and Kawasaki, M. and Nagaosa, N. and Tokura, Y.},
  year = 2016,
  month = jun,
  journal = {Nature Phys},
  volume = {12},
  number = {6},
  pages = {555--559},
  publisher = {Nature Publishing Group},
  issn = {1745-2481},
  doi = {10.1038/nphys3671},
  urldate = {2025-06-04},
  copyright = {2016 Springer Nature Limited},
  langid = {english}
}

@article{yuQuantizedAnomalousHall2010,
  title = {Quantized {{Anomalous Hall Effect}} in {{Magnetic Topological Insulators}}},
  author = {Yu, Rui and Zhang, Wei and Zhang, Hai-Jun and Zhang, Shou-Cheng and Dai, Xi and Fang, Zhong},
  year = 2010,
  month = jul,
  journal = {Science},
  volume = {329},
  number = {5987},
  pages = {61--64},
  publisher = {American Association for the Advancement of Science},
  doi = {10.1126/science.1187485},
  urldate = {2024-05-03}
}

\end{document}